# On the possibility of habitable Trojan planets in binary star systems

Richard Schwarz, Barbara Funk, Ákos Bazsó

Department of Astrophysics, University of Vienna, Austria

richard.schwarz@univie.ac.at

## Abstract

Approximately 60 percent of all stars in the solar neighbourhood (up to 80 percent in our Milky Way) are members of binary or multiple star systems. This fact led to the speculations that many more planets may exist in binary systems than are currently known. To estimate the habitability of exoplanetary systems, we have to define the so-called habitable zone (HZ). The HZ is defined as a region around a star where a planet would receive enough radiation to maintain liquid water on its surface and to be able to build a stable atmosphere. We search for new dynamical configurations - where planets may stay in stable orbits – to increase the probability to find a planet like the Earth. Therefore we investigated five candidates and found that two systems (HD 41004 and HD 196885) which have small stable regions.

## Introduction

Today we know about 3000 exoplanets thereof 100 exoplanets are found in binary star systems and approximately 20 exoplanets in multiple (triple) star systems.
The data and statistics of all planets are collected in the Exoplanet-catalogue maintained by J. Schneider[1] at (see Schneider et al. 2011) whereas the binary- and multiple-star systems can be found at binary catalogue of exoplanets maintained by R. Schwarz[2] (see Schwarz et al. 2016).

---

[1] http://exoplanet.eu
[2] http://www.univie.ac.at/adg/schwarz/multiple.html

When we speak about the possibility of habitable Trojan planets in binary star systems we have to answer two questions:

- **Why do we search for exoplanets in binaries?** Because many stars in the solar Neighbourhood are members of binary or multiple star systems. This fact leads to the conclusion that the knowledge of the dynamics of exoplanets in binary systems will be important in the future.
- **Why do we search for Trojan planets?** Because we are searching for new dynamical configurations where planets may stay in stable orbits. The more planets we will find the higher the possibility will be, to find a planet like the EARTH!

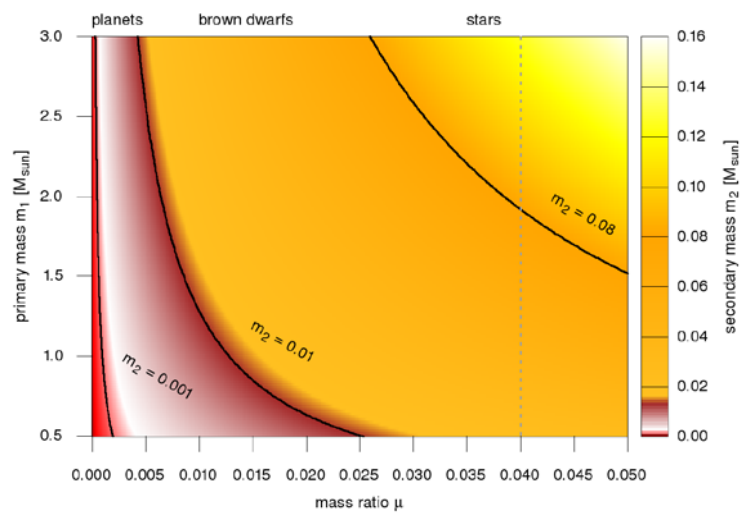

**Fig. 1** Dependence of the secondary star mass $m_2$ on the mass ratio $\mu$ and the mass of the primary star $m_1$ (main sequence stars between $0.5 < M_{sun} < 3$).

Jupiter's Trojan asteroids are in a 1:1 mean motion resonance with Jupiter, librating around the $L_4$ or $L_5$ points of the Sun–Jupiter system, preceding ($L_4$) or following ($L_5$) Jupiter at about 60°. Today we know approximately 5100 Jovian, 8 Neptunian and 5 Martian Trojans, but also one Uranus Trojan (2011QF99) and one Earth Trojan (2010TK7) were found.

It is known that in the planar, circular restricted three-body problem the equilibrium points $L_4$ or $L_5$ are stable for mass ratios up to $\mu = 0.04$; this stability limit is marked in Fig.1 as dashed vertical line. The mass ratio $\mu$ is defined as $\mu = m_2 / (m_1 + m_2)$ at which $m_1$ and $m_2$ being the masses of the primary and secondary bodies. So it seems that binaries may not have a population of Trojan asteroids or a Trojan planet, because the masses of the stars involved are comparable. However in a former study (Schwarz et al. 2009) we could show that binaries can have such mass ratios.

In Fig. 1 we visualize the equation $\mu = m_2 / (m_1 + m_2)$ and show the dependence of the mass ratio to the masses of the primary and secondary body. In this figure we show 3 limits: the first border shows a secondary body with $m_2$ for 0.001 $M_{sun}$ (given in solar masses) which corresponds to 1 Jupiter mass. The second border presents the transition from planetary bodies to substellar companions (at approximately $m_2$=0.013 $M_{sun}$ we have deuterium fusion) whereas the third limit show stellar objects. Finally we want to mention that the stability limit is larger for higher inclinations of the Trojan planet and can reach values up to $\mu = 0.048$ for i = 40°, shown in the work of Schwarz et.al (2012).

The dynamical behaviour and stability of planets in binaries are a very interesting and complex problem. In general one can distinguish different configurations in binary systems (Fig. 2):

**S-Type:** a planet orbits one of the two stars
**P-Type:** a planet orbits the entire binary
**T-Type:** a planet may orbit close to one of the equilibrium points $L_4$ and $L_5$ of the secondary star, which is shown in the work of Schwarz et al. (2009; 2012).

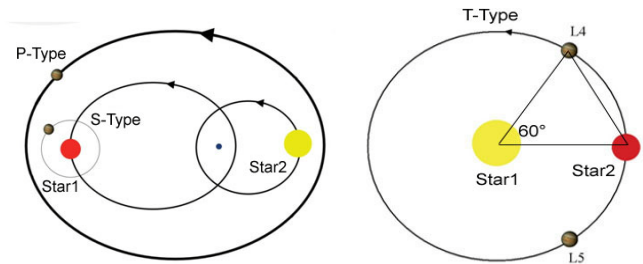

**Fig. 2.** Illustration of the different configurations in a binary system with one planet. The left side depicts the S and P-Type configuration whereas the right side show the Trojan configuration.

The question of habitability in binary star systems cannot be easily answered, because of the perturbation of the second star, whereas in most cases the additionally radiation can be neglected.

The search for Earth-like planets also raises the question of their habitability, and the so-called habitable zone (HZ) has been defined (e.g. Kasting, Whitmire & Reynolds 1993, Kaltenegger & Sasselov 2011).

This makes the definition of the HZ a highly interdisciplinary topic, including e.g. mass and atmosphere of the planet, properties of the host star like activity and radiation (e.g. Lammer et al. 2009; 2012) and also the dynamical stability of all involved bodies especially the eccentricity, since too large eccentricities can also lead to an unhabitable planet (Williams & Pollard 2002).

The dynamical stability is of course of special interest in binaries, since the secondary star could induce eccentricity to the planet. Thus Eggl et al. (2012) investigated the implications of stellar companions with different spectral types on the insolation a terrestrial planet receives orbiting a Sun-like primary. Their results suggest a strong dependence of permanent habitability on the binary's eccentricity, as well as a possible extension of habitable zones toward the secondary in close binary systems.

In general the Trojan configuration (Fig. 2 right graph) is interesting in systems where a gas giant moves mainly or permanently in the HZ, because the perturbation of the gas giant would push other planets (Earth-like planets) out of the HZ. Therefore we focused on the dynamical stability of binary star systems where a gas giant moves in the HZ of one of the two stars (S-Type configuration) and a Trojan planet (like the Jupiter Trojan asteroids in the solar system) may move around the Lagrangian equilibrium points $L_4$ or $L_5$ of the gas giant; as presented in Fig. 2 right graph.

## Model and methods

We studied the spatial restricted four-body problem (SR4BP) with three massive bodies ($m_1$ = star 1, $m_2$ = star 2 and $m_{GG}$ = gas giant) and a fourth massless body ($m_{Tro}$ = Trojan) moving under their gravitational influence (presented in Fig.3) as used in the work of Schwarz et al. (2009) and Papadouris & Papadakis (2014). We have regarded all the celestial bodies involved as point masses and integrated equations of motion for an integration time up to $T_c = 10^6$ periods of the gas giant, for the stability maps. We used the Lie-method with an automatic stepsize control to solve the equations of motion (Hanslmeier & Dvorak 1984, Lichtenegger 1984 and Eggl & Dvorak 2010).

For the analysis of stability we used the method of the maximum eccentricity $e_{max}$. The $e_{max}$ method uses as an indication of stability a straightforward check based on the maximum value of the Trojan's eccentricity reached during the total integration time. If the Trojan's orbit becomes hyperbolic ($e_{max} > 1$) the system is considered to be unstable. In former works by Dvorak et al. (2004) and Schwarz et al. (2007) we could show that the comparison of the results of the $e_{max}$ with the

Lyapunov characteristic indicator (LCI) and the Fast Lyapunov Indicator (FLI)[3] are in a good agreement.

To analyse the orbital behaviour of the massless bodies $m_{Tro}$ (e.g. tadpole, horseshoe or satellite), we used the amplitude of libration, as done in the work of Freistetter (2006) or Schwarz & Dvorak (2012).

## Candidate systems for possible Trojan planets in the HZ

Habitability and HZ depend on the stellar flux at the planet's location as well as the planet's atmospheric composition. Therefore Kaltenegger & Haghighipour (2013) calculated the boundaries of the habitable zone (HZ) of planet-hosting S-type binary star systems.

In the work of Eggl et. al. (2012) and Funk et al. (2015) they took into account radiative and dynamical effects. In our work we searched for binary systems candidates were gas giant moves partly or mainly in the HZ, based on the studies of Funk et al. (2015) and Kaltenegger & Haghighipour (2013).

We found five binary star systems 55Cnc, HD 41004, HD125612, HD 196885 and Ups And, where a gas giant moves mainly or partly in the HZ of one star and investigated the dynamical stability of a possible Trojan planet. However, our studies showed that only two binary systems can have stable Trojan planets at HD 41004Ab and HD 196885 Ab.

In a next step we show for this two systems in more detail what fraction of an orbit of the giant planet may stay in the HZ. In case of the work of Funk we take the averaged habitable zone (AHZ) which includes all in the article used configurations and gives the planet's time-averaged effective insolation. In addition we compare these results with the work of Kaltenegger & Haghighipour (2013) by using the empirical HZ (EHZ) including the insolation of both stars. The results are shown in table 1, where ΔAHZ and ΔEHZ represent the percentage how much of the planet's orbit stays in the AHZ/EHZ. We can conclude that HD 196885 is a more promising candidate for an additional Earth-like Trojan planet in case of habitability.

**Table 1:** The Estimates of the boundaries of the binary AHZ (Funk et al. 2015) and the empirical HZ (in AU). (*) In case of HD 41004 Ab there are no estimates for the EHZ, but we used the general values for a K1V star (α Cen B) which corresponds to HD 41004. The data can be found in Kaltenegger & Haghighipour (2013) in table 3.

| Name | inner AHZ | outer AHZ | Δ AHZ | inner EHZ | outer EHZ | Δ EHZ |
|---|---|---|---|---|---|---|
| HD 196885 Ab | 1.99 | 3.49 | 86 % | 1.137 | 2.62 | 60% |
| HD 41004 Ab | 0.77 | 1.39 | 33 % | 0.542* | 1.315* | 27%* |

---

[3] The LCI and the FLI are well-known chaos indicators.

# Results

In a former work of Schwarz et al. (2009a) we investigated the size of the stable region around the Lagrangian point $L_4/L_5$ in the planar R4BP. We found out that the stability regions of the two equilibrium points are similar. In addition we could show with an empirical formula how the size of the stable region of the Lagrangian points correlates with the mass ratio and the eccentricity. Based on that work we chose the following initial conditions as given in the next paragraph.

The system HD 41004 is composed of a K1 V primary (HD 41004 A) and a M2 V secondary (HD 41004 B) in a distance of 20 AU (more details shown in Table 1). First investigations concerning the stability in the region of the HZ of the system HD 41004 were done by Pilat-Lohinger (2005). Another investigation was done for HD 41004 and HD 196885 by Funk et al. (2015). We concentrated our investigations on the possibilities that a Trojan planet can move close to the orbit of a giant planet in a binary star system. Because of uncertainties of the orbital elements of these two systems we decided to enlarge the parameter space, in that way, that we varied the initial eccentricities from $e_{Tro}=0$ up to $e_{Tro}=0.9$ (with a grid size of $\Delta e=0.01$)[4] as well as for the gas giant and in addition we varied the orbital inclination up to $i_{Tro}=50°$ (with a grid size of $\Delta i=10°$). We always started in the Lagrangian equilibrium point $L_4$ with a choice of the argument of pericenter $\omega_{Tro}=60°$ (mean anomaly $M_{Tro}=0°$). The initial conditions of the two systems (HD 41004 and HD 196885 as given in table 2) which are unknown from the observations were set to zero as well as mean anomaly and inclinations of $m_2$ and $m_3$.

The stability maps for HD 41004 Ab are presented in Fig. 4 depending on the initial eccentricity of the Trojan planet $e_{Tro}$ and the eccentricity of the gas giant $e_{GG}$ for different initial inclinations ($i_{Tro}$) of the Trojan planet. Every stability map corresponds to 8100 orbits. We reduced the initial condition of the stability map $e_{Tro}=0-0.5$ and $e_{GG}=0-0.5$, because of the small stable region.
The 6 stability maps in Fig. 4 represent the different initial inclinations of the Trojan planet (upper left graph $i_{Tro}=0°$, upper right $i_{Tro}=10°$, middle left $i_{Tro}=20°$, middle right $i_{Tro}=30°$, lower left $i_{Tro}=40°$ and lower right graph $i_{Tro}=50°$). It is visible that we have a maximum of stable orbits for $i_{Tro}=10°$ and $i_{Tro}=40°$. This is also shown in table 3 where we summarised our results by counting the number of stable orbits for all stability maps. The stability analysis showed in addition that the initial eccentricity $e_{Tro}$ and $e_{GG}$ should not exceed 0.25. For higher inclinations ($i_{Tro}=20°$ to $i_{Tro}=50°$) the eccentricity $e_{Tro}$ can reach higher values than that of the giant planet $e_{GG}$. Former studies in the R3BP (Schwarz et al. 2014b) showed in contrary that the stability decreases monotonic with the inclination. This study showed a difference between $i=10°$ and $i=50°$ (shown in table 3), because we used not the same dynamical model (R3BP vs. SR4BP) and other initial conditions; e.g. different masses of $m_1$ and $m_2$. In case of high inclinations (50°) secondary resonances are close to the calculated initial conditions as shown in the work of Schwarz et al. (2014a).

---

[4] Every stability map corresponds to 8100 orbits. We reduced the initial conditions of the stability map $e_{Tro}=0-0.5$ and $e_{GG}=0-0.5$, because of the small stable region.

The system HD 196885 consists of a F8 V star with a mass of 1.3 $M_{Sun}$, a M1 V companion and a planet orbiting the primary. Chauvin et al. (2011) considered astrometric data as well as all data from past RV surveys to constrain the physical and orbital properties of the HD 196885 system and could find a unique best solution, which is summarized in table 2. Furthermore Chauvin et al. (2011) investigated the stability of the system and showed that the system is more stable in a high mutual inclination configuration. The results of HD 196885 are not shown, but are summarised in table. 3 like for HD 41004. The number of stable orbits is smaller than that of HD 41004, because the mass and the eccentricity of the second star is larger and the planet has a larger distance to the primary star, which perturbs the giant planet more than in the binary system HD 41004. We can conclude that for the system HD 196885 there are only stable orbits up to inclinations of $i_{Tro}=30°$.

## Conclusions

Today we know approximately 70 exoplanets in binary star systems thereof in 5 systems a giant planet move in the HZ.
In our work we made a stability analysis to show if an additional planet may move in the HZ of a binary star system. Therefore we investigated the Trojan configuration where a planet can move in the same orbit of a giant planet, but moving either close to 60° ahead or 60° behind the giant planet.
We focused our stability analysis on the S-Type configuration with possible Trojan planets. Our results showed that 2 of the 5 binary star systems may harbor stable Trojan planets, namely HD 41004 and HD 196885. However, HD 41004 is the most promising candidate, because the initial eccentricity (up to $e_{Tro}=e_{GG}=0.25$) and inclinations (up to $i_{Tro}=50°$) are higher than that of HD 196885 ($e_{Tro}=e_{GG}=0.2$; inclinations up to $i_{Tro}=30°$).
From the stability maps (Fig. 4) we can conclude that if the observed initial eccentricities for the gas giants (HD 41004 Ab and HD196885 Ab) are correct then stable Trojan planets are not possible. Even in case of habitability these systems are not promising, because the orbit of the Trojan planet stays only partly in the HZ.
   Nevertheless, the Trojan configuration is interesting when a giant planet moves in the HZ, because an additional planet (Earth-like planet) cannot move in the same region in a stable orbit. The numerical investigation showed that a Trojan planet can move in stable orbits in binary star systems. Our future aim will be to investigate transit light curves to detect Trojan planets in binary star systems.

**Acknowledgments:** R. Schwarz, B. Funk and A. Bazsó want to acknowledge the support by the Austrian FWF project P23810-N16.

**Table 2:** The orbital parameters of the two known tight binary systems, hosting an exoplanet. Given are the Name, semi-major axis (a), eccentricity (e), inclination (i), argument of perihelion (ω), longitude of the ascending node (Ω), the mass and the spectral type. In addition we show the known statistical errors. The parameters which are not given in the table are set to zero.

| Name | a [AU] | e | ω [deg] | Ω [deg] | mass | Spec. | Ref. |
|---|---|---|---|---|---|---|---|
| HD 196885 A | - | - | - | - | 1.3 $M_{Sun}$ | F8V | Chauvin et al. 2011 |
| HD 196885 B | 21 | 0.42 | 118.1 | 79.8 | 0.45 $M_{Sun}$ | M1V | Chauvin et al. 2011 |
|  | ± 0.86 | ± 0.03 | ± 3.1 | ± 0.1 | ± 0.1 |  |  |
| HD 196885 Ab | 2.6 | 0.48 | 93.2 | - | 2.98 $M_{Jup}$ | - | Chauvin et al. 2011 |
|  | ± 0.1 | ± 0.02 | ± 3.0 | - | ± 0.05 |  |  |
| HD 41004 A | - | - | - | - | 0.7 $M_{Sun}$ | K1V | Santos et al. 2002 |
| HD 41004 B | 20 | 0.4 | - | - | 0.42 $M_{Sun}$ | M2V | Roell et al. 2012, Zucker et al 2004 |
| HD 41004 Ab | 1.6 | 0.39 | 97 | - | 2.54 $M_{Jup}$ | - | Roell et al. 2012, Zucker et al 2004 |
|  | - | ± 0.17 | ± 31 | - | ± 0.74 |  |  |
| HD 41004 Bb | 0.0177 | 0.081 | 178.5 | - | 18.37 $M_{Jup}$ | - | Zucker et al 2004, Santos et al. 2002 |
|  |  | ± 0.01 | ± 7.8 | - | ± 0.22 | - |  |

**Table 3:** The number of stable orbits for each stability map for the different initial orbital inclination from i=0° up to i=50°.

| HD 41004 Ab | | HD 196885 Ab | |
|---|---|---|---|
| initial inclination [deg] | number of stable orbits in % | Initial inclination [deg] | number of stable orbits in % |
| 0 | 10.11 | 0 | 7.8 |
| 10 | 12.11 | 10 | 6.7 |
| 20 | 10.34 | 20 | 10.2 |
| 30 | 5.11 | 30 | 11.7 |
| 40 | 12.46 | 40 | 0.0 |
| 50 | 4.6 | 50 | 0.3 |

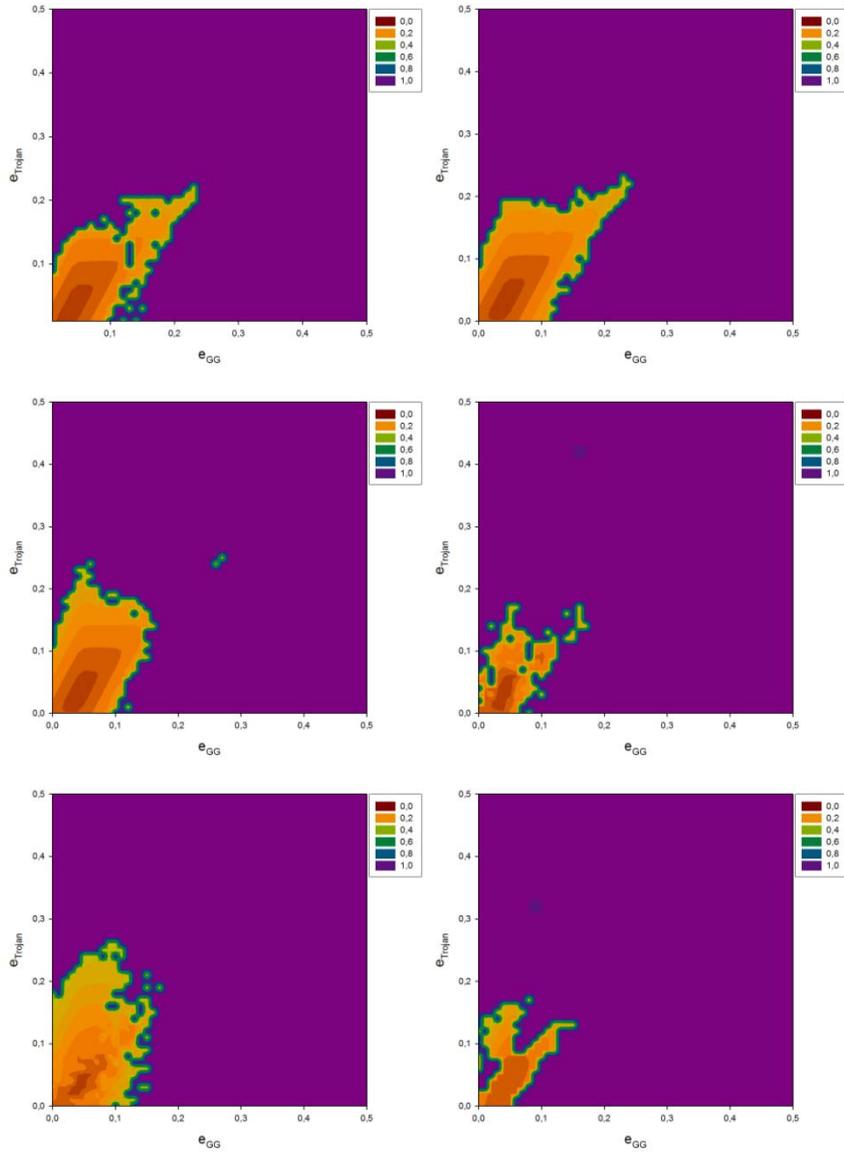

**Fig. 4.** The stability map of the system HD 41004 Ab, with 6 graphs: upper left graph $i_{Tro}=0°$, upper right $i_{Tro}=10°$, middle left $i_{Tro}=20°$, middle right $i_{Tro}=30°$, lower left $i_{Tro}=40°$ and lower right graph $i_{Tro}=50°$. The x-axis of the diagram depicts the initial eccentricity of the gas giant ($e_{GG}$) whereas the y-axis depicts the initial eccentricity of the Trojan ($e_{Tro}$). The colour-scale presents the values $e_{max}$, the red and orange colour depicts stable orbits and the violet shows unstable orbits. Every stability map corresponds to 8100 orbits.